\documentclass{iaus}
\usepackage{graphicx}
\include{epsf}

\def\simless{\mathbin{\lower 3pt\hbox
   {$\rlap{\raise 5pt\hbox{$\char'074$}}\mathchar"7218$}}}
\def\simgreat{\mathbin{\lower 3pt\hbox
   {$\rlap{\raise 5pt\hbox{$\char'076$}}\mathchar"7218$}}}   

\title[Constraints from strong lensing] 
{Constraints on dark and visible mass in galaxies from strong
gravitational lensing}

\author[Dye \& Warren]   
{Simon Dye$^1$\thanks{email: s.dye@astro.cf.ac.uk} \&
Steve Warren$^2$}

\affiliation{$^1$Cardiff School of Physics and Astronomy, Cardiff 
University, Queens Buildings, The Parade, Cardiff CF24 3AA, U.K. \\
$^2$Astrophysics Group, Imperial College London, Blackett Laboratory, 
Prince Consort Road, London SW7 2AZ, U.K. \\
}

\pubyear{2007}
\volume{244}  
\jname{Dark Galaxies and Lost Baryons}
\editors{}

\begin{document}

\maketitle

\begin{abstract}
We give a non-exhaustive review of the use of strong gravitational
lensing in placing constraints on the quantity of dark and visible
mass in galaxies. We discuss development of the methodology and
summarise some recent results.
\end{abstract}

\section{Introduction}

Quantifying the distribution of mass in the Universe is crucial to
understanding the physics that governs the formation of structure.  It
is therefore not surprising that measurement of the morphology and
fraction of dark matter in gravitationally bound systems has been a
common goal of countless studies to date. In particular, the
measurement of galaxy and cluster mass profiles has proven a powerful
observational probe for testing the cold dark matter (CDM) model.

Many different techniques have been applied to try to achieve such a
goal.  The methods used for measuring mass profiles generally
fall within one of four categories: dynamics, X-ray measurements, the
SZ effect and gravitational lensing. Each has its own set of
limitations and pitfalls but since these tend to be specific to
the approach used, the methods are largely complementary.

In the case of dynamical observations, complications in interpreting
results can arise from incorrect assumptions regarding orbital
motions, whether the system is relaxed and in radio data, the effect
of beam smearing on the shapes of mass profiles at small scales. X-ray
techniques relate the temperature of hot gas to the gravitational
potential assuming isothermal conditions apply. This limits the
technique to systems which are relaxed and in isothermal equilibrium
-- departures from these conditions result in a biased mass
profile. The scattering of cosmic background radiation by hot gas
exploited by the SZ method is potentially very powerful, being
detectable up to high redshifts, but with current technology is very
challenging and yields mass profiles with relatively poor
signal-to-noise.

Gravitational lensing has for some time now provided an attractive
alternative means of measuring mass profiles without the difficulties
associated with the other methods.  Primarily, this is motivated by
the simple fact that the deflection angle of a photon passing a
massive object is independent of the dynamical state of the mass
within the object. A limitation of lensing is that it is sensitive to
all mass along the line of sight, although foreground/background
alignments of mass occur much less frequently in galaxies than they
do on larger scales such as clusters.

A growing number of studies have focused on mass density profiles in
the centres of galaxies, comparing observations with the predictions
of CDM. \cite{nfw96} first proposed an analytic approximation to
describe the mass density profiles of halos in their CDM
simulations. At small radii (i.e., much smaller than a particular
scale radius), the mass density of this so called `NFW profile' scales
as $\rho(r) \propto r^{-1}$. Later simulations by \cite[Moore et
al. (1998,1999)]{moore98} indicated a steeper inner slope. This gave
rise to the `generalised' NFW (gNFW) profile which at radii smaller
than the scale radius follows $\rho(r) \propto r^{-\beta}$ with
values of $\beta$ around 1.4 to 1.5.  The most recent simulations
boasting a significantly higher resolution now converge on a slope
somewhere in the range $1.0\simless \beta \simless 1.2$
(\cite[Navarro et al. 2004]{navarro04}; \cite[Diemand et
al. 2005]{diemand05}).

In recent years, the slope of the inner mass profile has become a
subject of much contention with the results of simulations often being
very discrepant with observations. A complication in comparing slopes
from pure dark matter simulations with real measurements is that
baryons affect the morphology of the halo in a non-trivial way. The
canonical view is that baryons cause the halo to contract, steepening
its inner slope, and several prescriptions have been given to attempt
to describe this (for example, the adiabatic contraction model of
\cite[Blumenthal et al. 1986]{blumenthal86} or more recently that of
\cite[Gnedin et al. 2004]{gnedin04}). Regardless of the details,
the simple fact that contraction occurs means that inner halo slopes
determined from observations are actually an upper limit to the slope
the halo would have in the absence of baryons.

Estimates of the inner slope from dynamics are typically the most
discrepant with predictions. Several groups measuring rotation curves
of low surface brightness galaxies (LSBs -- believed to have a high
dark matter fraction and therefore be minimally affected by baryons)
find a range of slopes; $0\simless
\beta \simless 1$ (\cite[de Blok et al. 2001; de Blok \& Bosma 2002;
Swaters et al. 2003; Spekkens \& Giovanelli 2005]
{deblok01,deblok02,swaters03,spekkens05}).
\cite{hayashi04} purported that this discrepancy could be reconciled
by directly comparing against rotation curves of simulated
halos. However, this was strongly rejected by \cite{deblok05} who
found that only one quarter of the 51 galaxies in the study of
\cite{hayashi04} were consistent with CDM. In the latest episode of
this ongoing debate, \cite{hayashi06} claim that non-circular motions
in simulated CDM halos arising from a triaxial potential can explain
the range of measured rotation curves seen in LSBs. 

The findings of gravitational lensing studies, in particular those of
strong lens systems where the inner slope can be measured more
precisely than with weak lens systems, are generally in better
agreement with the CDM prediction. Strong lens systems allow lens
mass profiles to be constrained by searching for the best fit to the
observed multiple image positions of a background
source (see, for example, the review by
\cite[Schneider, Kochanek \& Wambsganss 2006]{schneider06}). 
The lens model typically takes on a parameterised form
but non-parametric models have also been explored (see, for example,
the pixelised mass modelling by \cite[Saha \& Williams 1996]{saha97}
applied most recently to mapping substructure in three strong lens
systems by \cite[Saha, Williams \& Ferreras 2007]{saha07}).
\cite{sand02} enhanced the multiple image fitting technique by 
incorporating extra constraints from the velocity dispersion profile
of the lens, and this has since seen application to a number of
systems (\cite[Treu \& Koopmans 2002; Koopmans \& Treu 2003; Sand et
al. 2004]{treu02,koopmans03,sand04}).  Nevertheless, \cite{dal03} have
criticised these results, arguing that the tight constraints claimed
were driven by prior assumptions and that in general, more detailed
modelling is required.

\subsection{Arcs and Einstein rings: A history of the methodology}

If the background source has extended structure, multiple arc
images or Einstein rings are formed. In high--resolution data, the
images comprise a large number of resolution elements. Extended
sources therefore have the considerable advantage that they can
provide many more constraints on the lens mass profile compared to
images of point sources. A complete analysis of images of extended
sources requires modelling of the source surface brightness
distribution. The properties of both the source and the lens must
be adjusted to give the best fit to the observed ring. 

One approach to this problem, suggested by \cite{ks88}, uses the fact
that regions of the source that are multiply--imaged have the same
surface--brightness. For a trial mass distribution, the method traces
image pixels to the source plane where the counts in different image
pixels mapping to the same source pixel are compared. The solution for
the mass is obtained by minimizing the dispersion in the image pixel
counts for such multiply--imaged source pixels. \cite{k89}
successfully applied this approach to the inversion of the radio
Einstein ring MG1131+0456. The algorithm was refined by \cite{wkk95}
who applied it to the triply--imaged giant arc in the galaxy cluster
Cl 0024+1654.

The main shortcoming of this approach is that it does not deal with
the image point spread function (PSF). If PSF smearing of the image
(either instrumental or atmospheric) is significant, the light profile
of the source is not correctly recovered by backward tracing the
image, even if the mass distribution is exactly known. To deal with
the PSF, a forward approach is needed i.e., one chooses a model for the
source light profile (parameterised or pixelised), and a model for the
mass (parameterised or pixelised), forms the image, convolves it with
the PSF, and compares it to the actual image, adjusting the source and
lens models to minimise a merit function e.g. $\chi^2$.

An argument for choosing to parameterise rather than pixelise the
source light profile is that it forces the solution to be smooth.
Nevertheless, the source light profile may be complex, as in the cases
of MG1131+0456 and Cl\,0024+1654 cited previously.  A large number
of parameters might be required to provide a satisfactory
description. Without clues to the character of the source it is
extremely difficult to select the best parameterization i.e. the one
which provides a satisfactory fit with the smallest number of
parameters. In the most extreme example \cite{ty98} used 232
parameters to model the source light distribution of the galaxy lensed
by the cluster Cl0024+1654.

If the source light profile is complex it is natural to consider
pixelizing the source, i.e. the counts in each pixel is a free
parameter. This removes the difficulty in finding a good
parameterization for the source, and thereby avoids any bias in the
fitted mass profile resulting from a poor choice. On the other hand,
due to the deconvolution, and because the pixels are independent, the
solution can be noisy. It is possible to achieve a smooth pixelised
solution by adding a suitable `regularizing' term to the merit
function.  \cite{wkn96} apply this approach to the case of the radio
Einstein ring MG 1654+134. They use a maximum entropy approach i.e.,
the regularizing term to be minimised is the negative of the entropy.

\cite[Warren \& Dye (2003; hereafter WD03)]{warren03} introduced a new
technique, which simplified and clarified the problem in a number of
ways. Formally, the method is very similar to the maximum entropy
method but WD03 showed that for a fixed lens mass distribution, the
minimization of the merit function is a linear problem and as such can
be solved by matrix inversion. This linear step is nested inside a
usual minimisation over the non-linear parameter space of the lens
model, hence the technique was called the `semi-linear method'.  The
advantages are that the total parameter space is much reduced,
dramatically improving the speed of the minimisation and greatly
easing location of the global minimum. WD03 also showed that the
method vastly simplifies calculation of the source and lens
uncertainties.

The semi-linear method has been applied to several systems to date
(\cite[Treu \& Koopmans 2004; Treu et al. 2006; Koopmans et al.
2006]{treu04,treu06,koopmans06} -- see next section). The method has
also been enhanced by a number of authors. 
\cite[Dye \& Warren (2005; hereafter DW05)]{dye05} showed how the
source plane can be adaptively gridded to minimise the covariance
between pixels and maximise use of the information contained in the
ring image.  A Bayesian version of the semi-linear method was
developed by \cite{suyu06}. The improvement this brings to the
original method is that comparison of the fit given by different
families of lens models is accomplished in a more rigorous fashion
through the Bayesian evidence (see \cite[MacKay 1992]{mackay92}).  In
addition, the Bayesian version determines the optimal level of
regularisation. Another modification was made by
\cite[Dye et al. (2007; hereafter D07)]{dye07} who showed how the
method can be generalised to allow for multiple background sources,
each at a different redshift.  The most recent enhancement was by
\cite{barnabe07}. In this version, constraints provided by the ring
image are combined with observed stellar velocity moments under a
Bayesian framework to place even more stringent limits on the lens
mass profile.

\section{Recent studies}

Adhering to the subject of this review, in this section we limit our
discussion to recent studies of strong lensing by galaxies where the
modelling accounts for the separate contribution of dark and baryonic
matter. All studies mentioned here make the common assumption that the
baryonic material follows the observed distribution of light and that
this is embedded in a dark matter halo, typically modelled by a NFW or
gNFW profile.

\begin{figure}
\epsfxsize=120mm
{\hfill
\epsfbox{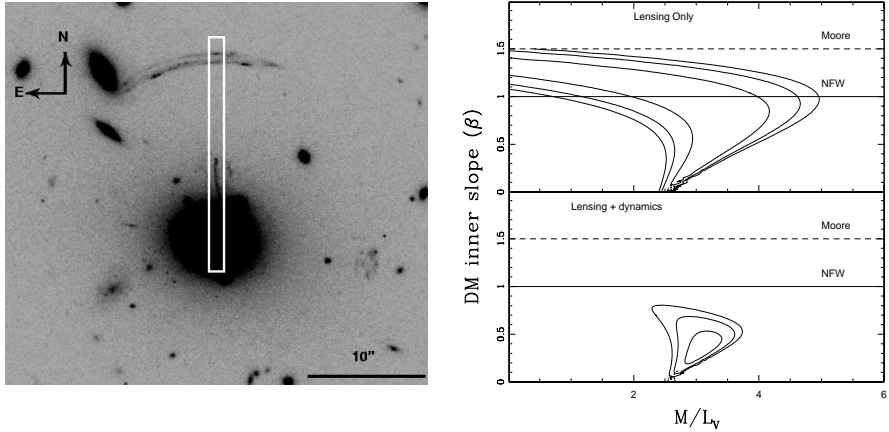}
\hfill}
\caption{Dual component modelling of the cluster MS2137-23 by
\cite{sand02}. The positions of the radial and tangential arc seen in
the image on the left are fit to give the constraints on the inner
dark matter halo slope and $V$ band stellar M/L shown at the top of
the plot on the right.  The lower plot includes the additional
constraints provided by the observed velocity dispersion profile. The
lens and source redshifts are respectively 0.31 and 1.50.}
\label{sand02_recon}
\end{figure}

\cite{sand02} fitted the positions of the observed radial and tangential
arcs with a dual component model for the brightest cluster galaxy in
MS2137-23. The authors showed how the inclusion of extra constraints
from the velocity dispersion profile allows much stronger limits to be
placed on the inner slope and stellar mass. The claim was made that
the cluster mass profile is completely inconsistent with CDM (see
Figure \ref{sand02_recon}), the halo having an inner slope of $\beta
\sim 0.4$. The same method and conclusions were also reached by
\cite{sand04}.  These results were criticised by \cite{dal03} who
argued that the constraints were artificially tightened due to
unrealistic assumptions regarding the galaxies' sphericity and scale
radius.

\begin{figure}
\epsfxsize=130mm
{\hfill
\epsfbox{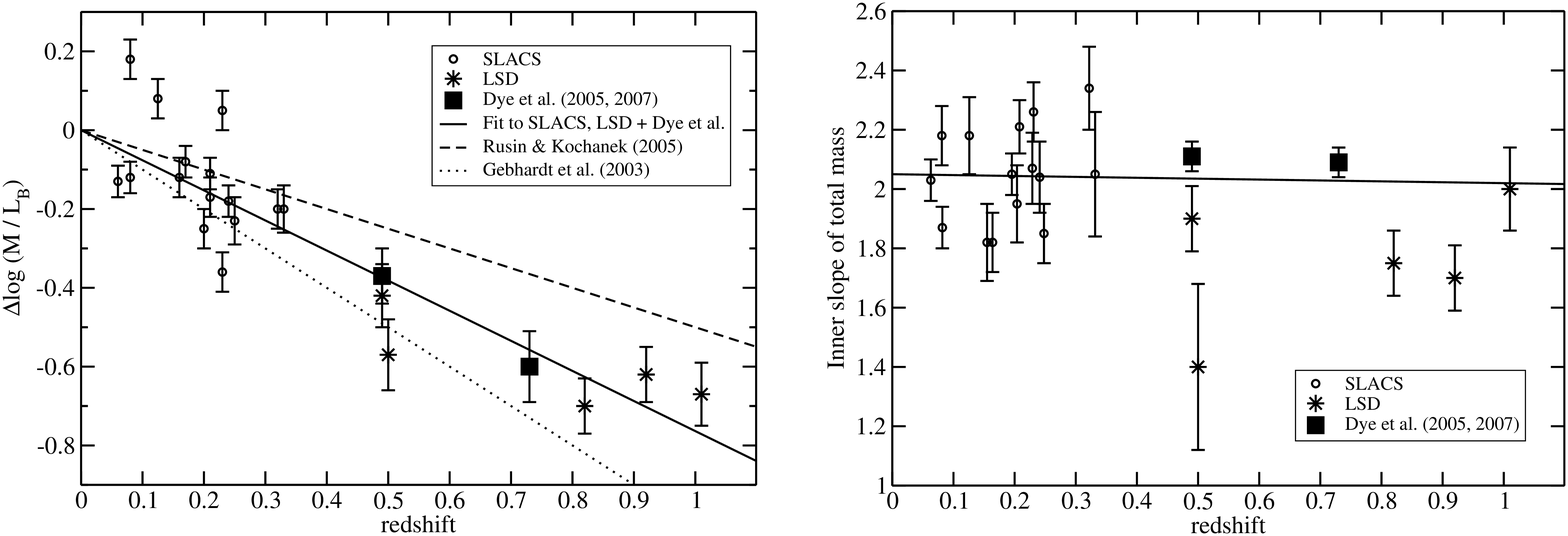}
\hfill}
\caption{{\em Left}: Evolution of the stellar M/L in lens galaxies
relative to local ellipticals. Data points are LSD (taken from
\cite[Treu \& Koopmans 2004]{treu04}), SLACS (taken from
\cite[Koopmans et al. 2006]{koopmans06}) and DW05, D07. The continuous
line is the best fit to all data points and has a gradient of
$-0.78\pm0.03$. The remaining two lines are two extremes taken from
the literature. The dashed line is from the lens sample of
\cite{rusin05} and the dotted line is from \cite{gebhardt03} for field
galaxies. {\em Right}: Evolution of the slope of the total (dark and
baryonic) mass. The best error-weighted straight line fit to all
points is $(2.05\pm0.05)-(0.03\pm0.11)z$, consistent with no
evolution.}
\label{evolution}
\end{figure}

In a similar study, \cite{treu04} modelled three strong galaxy lens
systems with a dual component model as part of the Lensing and Stellar
Dynamics project (LSD; \cite[Koopmans \& Treu 2002]{koopmans02}).
This study included central velocity dispersions of the lens galaxies
to tighten constraints on the modelling. A large scatter in the inner
halo slope was found, with a mean value of $\beta=1.3$. Similarly,
the fraction of projected dark matter contained within the Einstein
radius showed a large variation with values ranging between 0.37 and
0.72. A significant evolution of the $B$ band mass-to-light ratio
(M/L) was found with a slope of ${\rm d} \log (M_*/L_B)/{\rm
d}z=-0.75\pm0.17$ (see Figure \ref{evolution}). This study also
applied the semi-linear method, but only for the purpose of
reconstructing the extended source in two of the systems (i.e., the
method was not used to constrain the mass model).

\begin{figure}
\epsfxsize=120mm
{\hfill
\epsfbox{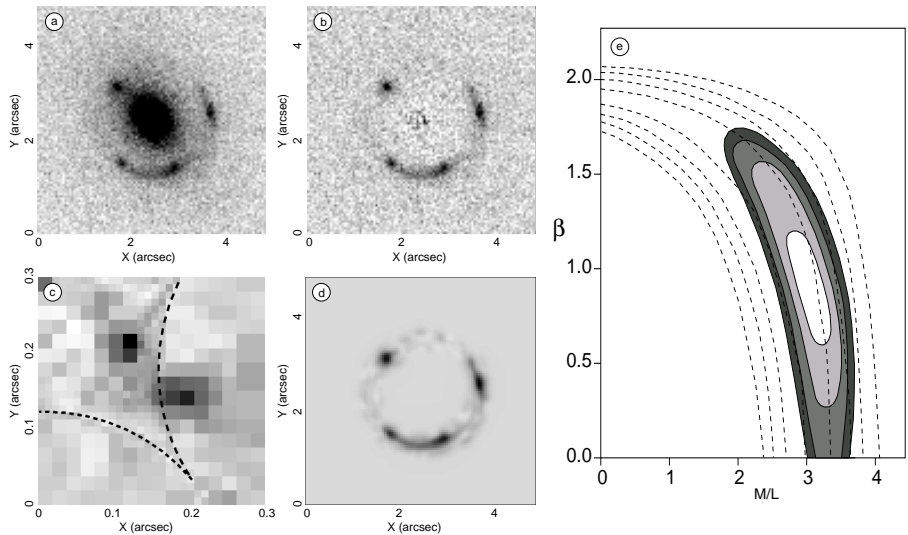}
\hfill}
\caption{Reconstruction of Einstein ring 0047-2808 from DW05.
a) Ring and galaxy lens ($z=0.49$ elliptical). b) Ring image after
subtracting Sersic profile. c) Reconstructed source (at $z=3.60$)
showing two distinct components. d) Image of the reconstructed source
using the best fit dual component model. e) 68\%, 95\%, 99\% \& 99.9\%
confidence regions on the best fit $B$ band stellar M/L and the inner
slope, $\beta$. The dashed lines are the same confidence regions from
the analysis by \cite{koopmans03} using the same data but only image
positional constraints.}
\label{dye05_recon}
\end{figure}

The first full application of the semi-linear method was that of DW05
to the Einstein ring system 0047-2808.  A summary of the results of
this work are shown in Figure \ref{dye05_recon}.  The study concluded
that a dual component model provides a better fit to the system than
any of the common single component models, very strongly ruling out a
model where the total mass follows the light. The best fit model
comprised a baryonic component with a $B$ band M/L of
$4.69^{+0.40}_{-0.69} \,h_{100} \,M_{\odot}/L_{B\odot}$ accounting for
$65\pm7\%$ of the projected mass within a mean ring radius of 1.16''
($\equiv 4.9h_{100}^{-1}$kpc. Deprojection gives a baryonic mass
fraction of $84\pm12\%$ within in a spherical volume enclosed by this
radius).

The analysis of DW05 was the first to constrain a stellar M/L
from a pure lensing analysis. This was a demonstration of the improved
constraints provided by all of the information contained in the
ring image, rather than just the positions of the principle images or
the ring radius. Panel (e) in Figure \ref{dye05_recon} compares the
constraints on the inner halo slope and stellar M/L with those of
\cite{koopmans03} who used positional information only (shown as dashed
lines).

\begin{figure}
\epsfxsize=130mm
{\hfill
\epsfbox{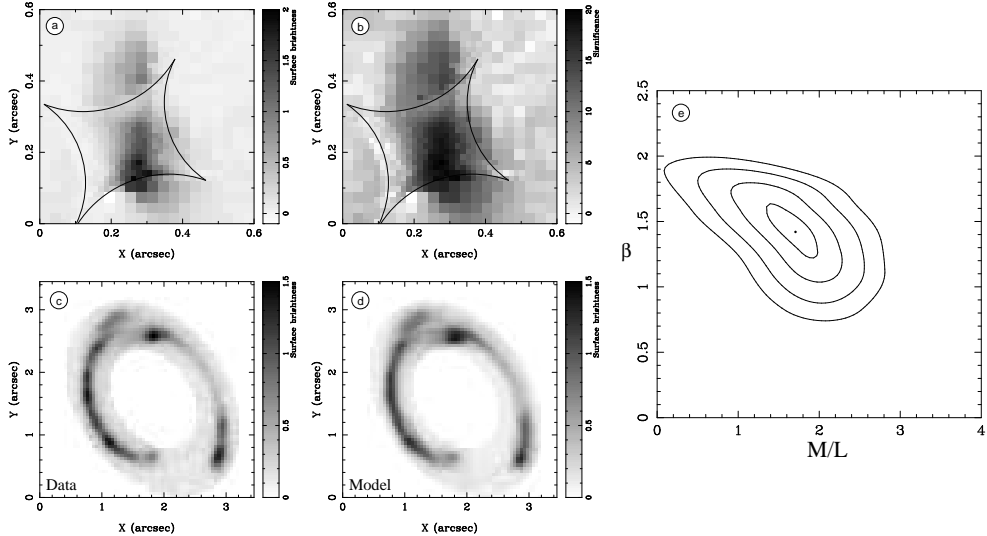}
\hfill}
\caption{Reconstruction of Einstein ring J2135$-$0102 (the `Cosmic
Eye') from D07.  a) Reconstructed source (at $z=3.07$) showing an
elongated brighter source straddling the caustic and a fainter source
just outside, b) The significance of the reconstructed source, c) The
observed ring image masked by an elliptical annulus, d) The image of
the reconstructed source formed using the best fit dual component
model, e) The 1, 2, 3 and 4$\sigma$ confidence contours on the $B$
band stellar M/L and the dark matter halo inner slope, $\beta$.}
\label{dye07_recon}
\end{figure}

The most recent application of the semi-linear method is that by D07
to the extraordinary Einstein ring system J2135$-$0102 (the 'Cosmic
Eye', \cite[Smail et al. 2007]{smail07}). The lens galaxy is an S0 at
$z=0.73$ and produces the incredible ring image of a background Lyman
break galaxy at $z=3.07$ shown in Figure \ref{dye07_recon}. The
ellipticity of the ring image is partly due to lensing by the
foreground cluster MACS J2135.2-0102 lying approximately $75''$ away
to the south at $z=0.33$. D07 modelled the system with a dual
component lens, finding that the dark matter halo has an inner slope
of $1.42^{+0.24}_{-0.22}$, consistent with CDM simulations after
allowing for baryon contraction.  The baryonic component was found to
have a mass-to-light ratio of $1.71^{+0.28}_{-0.38}$
M$_{\odot}$/L$_{B\odot}$ which when evolved to the present day is in
agreement with local ellipticals (see Figure \ref{evolution}). Within
the Einstein radius of $0.77''$ ($3.9 h_{100}^{-1}$kpc), the baryons
account for $(46\pm 11)$\% of the total projected lens mass.

The Sloan Lensing ACS (SLACS) Survey (\cite[Bolton et
al. 2006]{bolton06}) is a recently initiated HST snapshot imaging
programme for new galaxy-scale strong gravitational lenses. The new
lenses are identified in Sloan Digital Sky Survey (SDSS) spectra by
having emission lines at redshifts much higher than the targeted
galaxy.  These are followed up with shallow, single-orbit images
acquired with the ACS (now WFPC2). The survey is optimised for early
type lens galaxies and because it draws from SDSS spectra, is biased
towards relatively low redshift lenses ($z\simless 0.3$). SLACS is
rapidly increasing the number of known strong galaxy lens systems with
extended sources and at the time of writing, the project has
accumulated around 80 lenses (see Koopmans et al., these proceedings).

\cite{bolton06} presented the first 19 newly detected SLACS
lenses. The second SLACS paper by \cite{treu06} analysed the
photometric properties and velocity dispersions of 15 of these and
found that the galaxies are consistent with the parent SDSS sample but
belong to a subset of bright (and high velocity dispersion) early
types. Combined with the lens galaxies from the LSD project, the
evolution in the $B$ band M/L was measured to be ${\rm d} \log
(M_*/L_B)/{\rm d}z=-0.76\pm0.03$ (see Figure \ref{evolution}).  The
findings of DW05 and D07 are completely consistent
with this evolution (combining the data from these two studies with
the LSD and SLACS data gives an overall evolution of ${\rm d} \log
(M_*/L_B)/{\rm d}z=-0.78\pm0.03$). \cite{treu06} concluded that this
evolution is consistent with passive ageing of an old stellar
population formed at $z>2$ and that no more than 10\% of the stellar
mass in these systems can have formed due to secondary star
formation since $z=1$.

In the third SLACS paper, \cite{koopmans06} carried out a joint
lensing and dynamics analysis of the same 15 strong lenses as
\cite{treu06}. The work applied the semi-linear method to reconstruct
source surface brightness profiles, but carried out a separate
analysis, using only the central velocity dispersion and total mass
within the Einstein radius to determine the slope of the total (i.e.,
dark and baryonic) density profile. Combined with the findings of the
LSD project, the total slope was found to be non-evolving over the
redshift interval $0<z<1$ (see Figure \ref{evolution}). Adding the
total mass slopes found in DW05 and D07 to the LSD
and SLACS data points gives an error weighted straight line fit of
$(2.05\pm0.05)-(0.03\pm0.11)z$. Although at present the SLACS lenses
lack sufficient spatial resolution in their kinematical data to enable
measurement of the properties of just the halo component per galaxy,
\cite{koopmans06} calculate an average projected dark matter fraction
inside the Einstein radius of 0.25 with an rms of 0.22. This large
scatter, which will inevitably also be seen in the inner halo slope,
is surprising given the consistency of the total slope and its lack of
evolution. The mechanisms driving this 'bulge halo conspiracy' are
currently very unclear.

\begin{figure}
\epsfxsize=100mm
{\hfill
\epsfbox{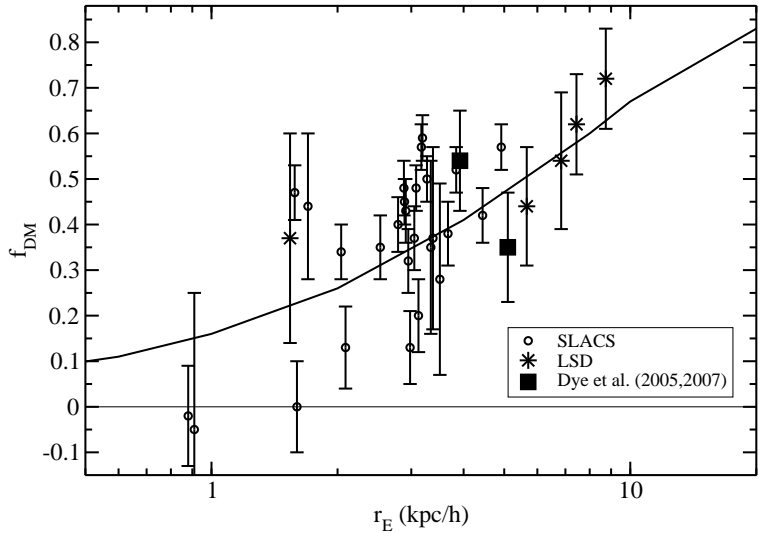}
\hfill}
\caption{Projected dark matter fraction within the Einstein radius as
a function of the Einstein radius for LSD, SLACS and Dye et
al. (2005,2007). The solid line is based on the average baryon + dark
matter halo model fit to the 22 SLACS lenses by \cite{gavazzi07}.}
\label{f_dm_vs_z}
\end{figure}

In the fourth and most recent SLACS paper, \cite{gavazzi07} have
determined projected dark matter fractions within the Einstein radius
per galaxy by combining the strong lensing analysis of the previous
papers with weak lensing measurements. In Figure \ref{f_dm_vs_z} we
add the LSD lenses and those of DW05 and D07 to the SLACS lenses
(including the new additions) of \cite{gavazzi07}. The figure shows that
the larger and more distant lenses of LSD, DW05 and D07 are consistent
with the average dark matter fractions of the more local SLACS lenses
(shown by the continuous line).

\cite{koopmans06} note their intention to address the lack of spatial
resolution in their dynamical observations of the SLACS lenses in
future work. This should enable a full dual component lens analysis
such that characteristics of the halo, in particular the inner slope,
can be determined on a galaxy-by-galaxy basis. In this way, one would
hope that insight into the bulge halo conspiracy could be gained.
However, in order to fully establish the relationship between the
morphology of the baryons and the morphology of the dark matter, its
evolution must be properly explored. Here, the SLACS study with its
relatively low redshift sample of lenses falls somewhat short and
studies which concentrate on more detailed modelling of higher
redshift systems (e.g., \cite[Treu \& Koopmans 2004]{treu04},
\cite[Dye \& Warren 2005]{dye05}, \cite[Dye et al. 2007]{dye07}) play
a vital role.

We recently carried out ACS High Resolution Channel (HRC) imaging of
three Einstein ring systems (HST proposal \#10563). The lenses are
early type galaxies spanning the redshift range $0.3<z<0.6$. Our
approach differs from the SLACS philosophy in two main ways. Firstly,
we obtain deeper images (e.g., one of our lenses received 12 orbits of
integration). In this way, we obtain a higher signal-to-noise thus
being able to carry out more detailed modelling, placing stronger
limits on lens parameters and being able to more strongly delineate
between different lens models. Secondly, our imaging is carried out
with the ACS's higher resolution HRC, unlike SLACS which uses
the Wide Field Camera (WFC) and now WFPC2. We have carried out
simulations which show that despite the lower throughput of the HRC
relative to the WFC, for the same number of orbits, the
semi-linear method performs slightly better on HRC data than on WFC
data in terms of both constraining the lens model and the source
reconstruction. Analysis of these lenses is in progress and when 
finished will complement the SLACS project by adding to the number
of higher redshift systems.

\begin{acknowledgments}
SD is supported by PPARC. We thank Ian Smail, Mark Swinbank, Harald
Ebeling and Alastair Edge for their crucial involvement in the work on
the Cosmic Eye.
\end{acknowledgments}


\begin{thebibliography}{}

\bibitem[Barnab\`{e} \& Koopmans (2007)]{barnabe07} Barnab\`{e}, M.
\& Koopmans, L. V. E., 2007, ApJ submitted, astro-ph/0701372

\bibitem[de Blok et al. (2001)]{deblok01} de Blok, W. J. G., McGaugh, S. S., 
Bosma, A., Rubin, V. C., 2001, ApJ, 552, 23

\bibitem[de Blok \& Bosma (2002)]{deblok02} de Blok, W. J. G.,  Bosma, A.,
2002, A\&A, 385, 816

\bibitem[de Blok (2005)]{deblok05} de Blok, W. J. G., 2005, ApJ, 634, 227

\bibitem[Blumenthal et al. (1986)]{blumenthal86} Blumenthal, G.R., Faber,
S.M, Flores, R., Primack, J.R., 1986, ApJ, 301, 27

\bibitem[Bolton et al. (2006)]{bolton06} Bolton, A. S., Burles, S.,
Koopmans, L. V. E., Treu, T., Moustakas, L. A., 2006, ApJ, 638, 703

\bibitem[Diemand et al. (2005)]{diemand05} Diemand, J., Zemp, M., 
Moore, B., Stadel, J., Carollo, C. M., 2005, MNRAS, 364, 665

\bibitem[Dalal \& Keeton (2003)]{dal03} Dalal, N. \& Keeton, C.R., 2003,
astro-ph/0312072

\bibitem[Dye et al. (2007)]{dye07} Dye, S., Smail, I., Swinbank, A. M., 
Ebeling, H., Edge, A. C., 2007, MNRAS, 379, 308, (D07)

\bibitem[Dye \& Warren (2005)]{dye05} Dye, S. \& Warren, S. J.,
2005, ApJ, 623, 31, (DW05)

\bibitem[Gavazzi et al. (2007)]{gavazzi07} Gavazzi, R., Treu, T., Rhodes,
J. D., Koopmans, L. V. E., Bolton, A. S., Burles, S., Massey, R. J.,
Moustakas, L. A., 2007, ApJ, in press, astro-ph/0701589

\bibitem[Gebhardt et al. (2003)]{gebhardt03} Gebhardt, K., et al., 2003,
ApJ, 597, 239

\bibitem[Gnedin et al. (2004)]{gnedin04} Gnedin, O. Y., Kravtsov, A. V.,
Klypin A. A., Nagai, D., 2004, ApJ, 616, 16

\bibitem[Hayashi et al. (2004)]{hayashi04} Hayashi, E., Navarro, J. F.,
Power, C., Jenkins, A., Frenk, C. S., White, S. D. M.,
Springel, V., Stadel, J., Quinn, T. R., 2004, MNRAS, 355, 794

\bibitem[Hayashi \& Navarro (2006)]{hayashi06} Hayashi, E. \&
Navarro, J. F., 2006, MNRAS, 373, 1117

\bibitem[Kayser \& Schramm (1988)]{ks88} Kayser, R. \& Schramm, T.,
1988, A\&A, 191, 39 

\bibitem[Kochanek et al. (1989)]{k89} Kochanek, C.S., Blandford R.D.,
Lawrence C.R., \& Narayan, R., 1989, MNRAS, 238, 43

\bibitem[Koopmans \& Treu (2002)]{koopmans02} Koopmans, L.V.E.
\& Treu, T., 2002, ApJ, 568, L5

\bibitem[Koopmans \& Treu (2003)]{koopmans03} Koopmans, L.V.E.
\& Treu, T., 2003, ApJ, 583, 606

\bibitem[Koopmans et al. (2006)]{koopmans06} Koopmans, L.V.E., Treu, T., 
Bolton, A. S., Burles, S., Moustakas, L. A., 2006, ApJ, 640, 662

\bibitem[MacKay (1992)]{mackay92} MacKay, D. J. C., 1992, Neural
Computation, 4, 415

\bibitem[Moore et al. (1998)]{moore98} Moore, B., Governato, F., Quinn,
T., Stadel, J., Lake, G., 1998, Apj, 499, L5

\bibitem[Moore et al. (1999)]{moore99} Moore, B., Quinn,
T., Governato, F., Stadel, J., Lake, G., 1999, MNRAS, 310, 1147

\bibitem[Navarro et al. (2004)]{navarro04} Navarro, J. F., Hayashi, E., 
Power, C., Jenkins, A., Frenk, C. S., White, S. D. M., Springel, V., Stadel,
J., Quinn, T. R., 2004, MNRAS, 349, 1039

\bibitem[Navarro, Frenk \& White (1996)]{nfw96} Navarro, J. F., Frenk, C. S.
\& White S. D. M., 1996, ApJ, 462, 563

\bibitem[Rusin \& Kochanek (2005)]{rusin05} Rusin, D. \& Kochanek, C. S.,
2005, ApJ, 623, 666

\bibitem[Saha \& Williams (1997)]{saha97} Saha, P., Williams, L. L. R.,
1997, MNRAS, 292, 148

\bibitem[Saha, Williams \& Ferreras (2007)]{saha07} 
Saha, P., Williams, L. L. R. \& Ferreras, I., 2007, ApJ, 663, 29

\bibitem[Sand et al. (2004)]{sand04} Sand, D. J., Treu, T., Smith, G. P., 
Ellis, R. S., 2004, ApJ, 604, 88

\bibitem[Sand, Treu \& Ellis (2002)]{sand02} Sand, D. J., Treu, T., \&
Ellis, R. S., 2002, ApJ, 574, L129

\bibitem[Schneider, Kochanek \& Wambsganss (2006)]{schneider06}
Schneider, P., Kochanek, C. S. \& Wambsganss, J., 2004, Part 2 of 
Gravitational Lensing: Strong, Weak \& Micro, Proceedings of the 33rd 
Saas-Fee Advanced Course, (eds. G. Meylan, P. Jetzer \& P. North)

\bibitem[Smail et al. (2007)]{smail07} Smail, I., Swinbank, A. M.,
Richard, J., Ebeling, H., Kneib, J. -P., Edge, A. C., Stark, D.,
Ellis, R. S., Dye, S., Smith, G. P., Mullis, C., 2007, ApJ, 654, 33

\bibitem[Spekkens \& Giovanelli (2005)]{spekkens05} Spekkens, K., 
Giovanelli, R. \& Haynes, M. P., 2005, AJ, 129, 2119

\bibitem[Suyu et al. (2006)]{suyu06} Suyu, S. H., Marshall, P. J.,
Hobson, M. P., Blandford, R. D., 2006, MNRAS, 371, 983

\bibitem[Swaters et al. (2003)]{swaters03} Swaters, R.A., Madore, B.F.,
van den Bosch, F. C., Balcells, M., 2003, ApJ, 583, 732

\bibitem[Treu \& Koopmans (2002)]{treu02} Treu, T. \& Koopmans, L. V. E.,
2002, ApJ, 575, 87

\bibitem[Treu \& Koopmans (2004)]{treu04} Treu, T. \& Koopmans, L. V. E.,
2004, ApJ, 611, 739
	
\bibitem[Treu et al. (2006)]{treu06} Treu, T., Koopmans, L. V. E.,
Bolton, A. S., Burles, S., Moustakas, L. A., 2006, ApJ, 640, 662

\bibitem[Tyson, Kochanski \& dell'Antonio (1998)]{ty98} Tyson, J.A., 
Kochanski, G.P., \& dell'Antonio, I.P., 1998, ApJ, 498, 107

\bibitem[Wallington, Kochanek \& Koo (1995)]{wkk95} Wallington, S., 
Kochanek, C. S., \& Koo, D., 1995, ApJ, 441, 58

\bibitem[Wallington, Kochanek \& Narayan (1996)]{wkn96} Wallington, S., 
Kochanek, C. S., \& Narayan, R., 1996, ApJ, 465, 64

\bibitem[Warren \& Dye (2003)]{warren03} Warren, S. J. \& Dye, S., 
2003, ApJ, 590, 673, (WD03)

\end{thebibliography}
\end{document}